# Scientific knowledge production of blockchain: A bibliometric and lexicometric review

**Yuan Li**, yuan.li7@univ-lyon2.fr
**Wilfrid Azan**, wilfrid.azan@univ-lyon2.fr

## Abstract:

While recent reviews of blockchain technology have focused on the latest developments in cryptocurrency and their derivative impacts, less attention has been given to analysing their knowledge paths and boundaries based on past research to guide their development. To address this gap, we conducted both a bibliometric study of 2525 articles and a lexicometric study of 123 articles. The bibliometric study provided holistic insights into the evolution and distribution of blockchain research, including influential researchers and countries, discipline composition, knowledge development trends, and emerging frontiers. The lexicometric study identified the boundary concept structure with a quantitative textual approach, extracting the strongest signifying epistemic communities. Our findings indicate that blockchain research draws from four major disciplines, making it a multidisciplinary field. With the increasing maturity and development of technological infrastructure, the application and management of blockchain become increasingly relevant issues. Our analysis suggests that blockchain can be considered more of a boundary object than a disruptive change from knowledge perspectives. Therefore, this paper proposes a comprehensive understanding of the development path and epistemic concepts of blockchain research.

## Introduction

In 2001, Iansiti et al. questioned the effect of the Internet on architectural design (MacCormack et al., 2001). Fifteen years later, Iansiti and Lakhani (2017) have echoed this question regarding the emergence of blockchain. As a new information and communication technology (ICT), blockchain's emerging role has been compared to that of the Internet in the early 1990s. Blockchain implements a verification system for digital identity and records transactions chronologically (Swan, 2015). Beyond information sharing, it raises awareness to secure valuable information and establishes trustworthy interconnectedness (Turk and Klinc, 2017). For Nærland et al. (2017), blockchain facilitates exchange processes and inter-organizational systems that contribute to the development of trusting relationships. Blockchain has been regarded as a better way in ICTs for collaborative relationship management as it can build and engage in a common process of value creation in a business network (Lumineau et al., 2021). One of the most enticing promises of blockchain is its ability to support infrastructure maintenance through a decentralized network (Beck et al., 2017). Decentralization is the symbolic feature of blockchain that could displace traditional intermediaries (Schulze et al., 2020). However, the extent to which blockchains achieve decentralization remains open to debate.

There has been considerable interest in blockchain research, prompting a need to retrospectively analyse this research domain (Miau & Yang, 2018; Dabbagh et al., 2019). Some application-based reviews suggest that blockchain technology, equipped with scalability and interoperability, can further link the digital and physical worlds (Casino et al., 2019; Belchior et al., 2021). However, other technical reviews have found that there is still insufficient evaluation of the effectiveness of blockchain's privacy and security (Yli-Huumo et al., 2016). They argue that blockchain research is still in its early stages and lacks theoretical grounding, methodological diversity, and empirically grounded work (Frizzo-Barker et al., 2020; Ziolkowski et al., 2020). Previous reviews have conducted systematic literature reviews on specific fields or single methodologies, such as healthcare, organizational collaboration, agriculture, tourism, and technological evolution (Jabbar et al., 2021; Hölbl et al., 2018; Bermeo-Almeida et al., 2018; Conoscenti et al., 2016; Fragnière et al., 2022). However, these reviews do not reveal the full dynamics and snapshots of theory development and conceptual classification. Leveraging the literature on knowledge management, it is worth noting that technology transfer processes have strategic implications for firms (Coadour et al., 2019).

This paper provides a comprehensive overview of the literature on blockchain management from an interdisciplinary perspective. It integrates bibliometric analysis to trace the development path and lexicometric analysis to highlight behavioural trends and classify epistemic communities (CEs). The study aims to identify the scientific knowledge and epistemic foundation of blockchain through these analyses and achieves the following three research objectives (ROs):

RO1: To outline the main paths of blockchain management based on bibliometric citations.

RO2: To extract the strongest conceptual structures by grouping the quantified lexical text.

RO3: To summarize the boundaries and guide the future development of blockchain based on the combination of bibliometric and lexicometric studies.

This article makes two key contributions. Firstly, it pays particular attention to the evolution of blockchain management from its inception to its current mainstream status. By analysing 2525 research articles, the collaborative networks made by researchers across multiple disciplines can be characterized. The bibliometric method over a time axis shows the frontiers of blockchain emerging (Xu et al., 2019), providing insights for future research through co-citation network, co-occurrence analysis, and burst detection. Secondly, the article attempts to evaluate blockchain as a boundary object using lexicometric analysis. Based on two architectures of boundary concept by Levina and Vaast (2005), the boundary characteristics of blockchain are identified based on cognitive and hierarchical models. Within the cross-study, they defined the boundary objects by deriving from practices of various fields and obtaining a clear picture by identifying common practices (Levina and Vaast, 2006).

In the following sections, Section 1 outlines the fundamental principles of the blockchain software and presents the methodologies used in our research. Sections 2 and 3 visualize the main research paths and concept categories, using CiteSpace to conduct cluster analysis, identify top-cited articles, and detect citation bursts for development pathfinding and hot issues. In addition, Alceste and IRaMuTeQ are used to identify the epistemic communities of blockchain. In Section 4, we thoroughly discuss blockchain management by combining bibliometric and lexicometric aspects. The final section concludes our findings and implications.

## 1. Methodology and Data

In this section, we introduce the function and mechanism of research. We use bibliometric and lexicometric approaches to map the literature on blockchain management. Our study explores blockchain research with integral chronological streams and knowledge flows (Budler et al., 2021), which is particularly useful when identifying boundaries and knowledge communities.

**1.1. Research Design**

The clustering methods used in bibliometrics can help to delineate sub-streams of networks at a macro-level (Snyder, 2019). On the other hand, the textual approach can quantify text to extract the most significant structures of lexical fields at a micro-level. The interpretive approach was used to analyse the data. Our research design is depicted in Figure 1.

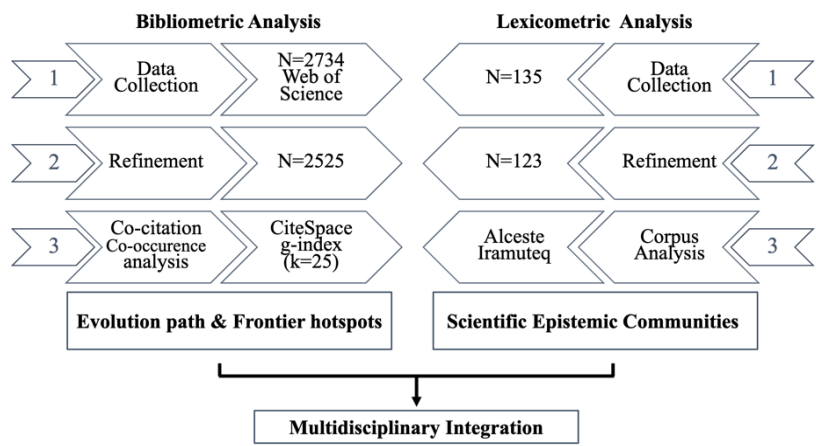

Figure 1. Research Design

Bibliometric study with CiteSpace: CiteSpace software, a Java-based application, is used to conduct bibliometric studies and present information visually by analysing quantitative reference data (Chen, 2006). By examining the co-citation data of academic research articles, it identifies correlations between knowledge flows and uncovers research trends (Guo et al., 2021). The availability of citations in publications reflects how the topic disseminates (Chen, 2004). Citation-based analysis has been widely adopted in scientific reviews to investigate knowledge clusters and research directions (Meng et al., 2021). The co-citation approach explores the correlation between literature cited together and can visualize the structure of the knowledge base and its evolution (Small, 1973).

This paper adopts three citation-based approaches, which have been used in prior studies (Xu et al., 2019; Chen et al., 2019; Wu et al., 2019; Guo et al., 2021): co-citation cluster analysis, co-occurrence analysis, and noun-term burst detection.

1. Network analysis: First, we provide a brief overview of the overall state of blockchain management, including distribution by country and discipline. Network analysis ensures that we have sufficient and high-quality data, maintaining the interdisciplinary standard of research.

2. Co-citation clustering and co-occurrence analyses: According to Chen et al. (2010), a research front consists of a cluster of co-cited core papers, as well as current source papers that cite one or more of these core papers. Timeline analysis can show how research hotspots develop: research history, current research, and forecasted future developments. The list of top-cited articles identifies the significant literature in blockchain research based on occurrence.

3. Noun-term burst detection refers to the sudden emergence or sharp increase of noun terms within a short period of time (Chen et al., 2010). It is a representative phenomenon in which the frequency of noun terms increases as the research frontier is pushed in a certain direction for a specified duration.

Lexicometric analysis is a method of quantifying texts to extract the most significant signifying structures of lexical fields. The concept of a Community of Experts (CE) is interesting to explore as it leads to new epistemological divisions. CEs are groups of individuals with expertise in a particular field, possessing recognized skills in legitimacy and authority, and capable of producing reliable academic content (McDermott, 2000; Cuddy, 2002). This concept has been extended by communities of exploration that add additional layers of knowledge to existing practices (Cross et al., 2004; Cohendet et al., 2001; Azan et al., 2017). In this study, we processed the data using the IRaMuTeQ software (Version 07 alpha 2) and subsequently reprocessed it using Alceste. IRaMuTeQ begins by lemmatizing the corpus and then analysing the textual data, retaining only the "active" forms for analysis. The software segments the text into text segments and performs a hierarchical descending classification, dividing the largest class into two groups and repeating the process. The classes are differentiated by the distribution of their vocabulary at each step. To restrict the classes to the forms present in both categories, the software performs two successive classifications. To facilitate the processing of a large volume of text and extract its underlying characteristics, we chose to use lexicometric analysis, which segments the corpus of texts into classes based on their vocabulary distribution. Newman (2006) defined the equations for detecting community structure as follows.

$$Q(C) = \sum_{i,j \in C} A_{ij} - \frac{d_i d_j}{2m}$$

Lexicometric analysis allows the differentiation of character strings into "raw words," which are listed in a dictionary and filtered based on relevant criteria. Moreover, it creates a numerical representation of the corpus that facilitates manipulation by other processing programs while preserving the contextual information of different units (Reinert, 1983). To identify each distinct word in the corpus, it is necessary to verify whether it already exists in the dictionary and add it if not. The solution employed in this study meets our speed requirements. Removing common words, which are already included in the dictionary, is crucial for reducing the size of the digital file associated with the transcribed corpus and the reading time. Additionally, the method focuses on detecting irregular words, complementing the algorithm by reducing the occurrence of the most common irregular forms and enhancing the analysis's clarity (Reinert, 1986).

The primary objective is to group raw forms that belong to the same "lexeme," regardless of their syntactic roles (Reinert, 1986). The software performs such groupings based on the selected lexicon and the markers associated with syntax, such as grammatical inflections and certain suffixes. By considering the broadest contexts to comprehend the semantic structure, we aim to minimize information loss and retain the maximum number of words. Grouping all applicable terms while discarding infrequently used words is crucial to accomplish the research goal. The program's purpose is to offer the user a tool to conduct groupings more efficiently and objectively (Reinert, 1983). To achieve this, it is necessary to group all words with a high frequency of semantic linkage by reducing them to their common roots and eliminating unlikely entries. Reinert refers to this as a morphological analysis strategy (Reinert, 1983).

### 1.2. Data

First, we clarify different databases collected in bibliometric and lexicometric studies. Then, we implement operational procedures to extract useful information. CiteSpace is appropriate for bibliometric analysis of citation records to explore research trends. Alceste and IRaMuTeQ conduct corpus analysis to extract accurate snapshots of concept communities. The bibliometric data covers the period from 2008 to 2021, and the lexicometric data ranged from 2008 to 2019.

In the CiteSpace analysis, we collected bibliometric data from the "Clarivate Analytics Web of Science" (WOS) Core Collection database, which is commonly considered the most comprehensive database for academic studies in bibliometric reviews (Zupic & Čater, 2015). We searched for the term 'blockchain management' as a topic and found 2734 results. After filtering out document types such as news, letters, and reports, we included only articles, proceeding papers, and books. We then downloaded and extracted full bibliographic records and cited references of the remaining 2525 results in plain text format. The database was last updated on May 2, 2021. We specified "k=25", which uses a modified G-index in each slice ($g^2 \leq k \sum_{i \leq g} C_i$, with each time slice set to one year). We can see from Table 1 that blockchain management has steadily garnered interest since 2018.

Table 1. Number of blockchain academic articles in WOS

| Publication Years | Record Count | % of 2525 |
|---|---|---|
| 2021 | 256 | 10.139 |
| 2020 | 997 | 39.485 |
| 2019 | 776 | 30.733 |
| 2018 | 367 | 14.535 |
| 2017 | 107 | 4.238 |
| 2016 | 17 | 0.673 |
| Before 2016 | 5 | 0.198 |
| Total | 2525 | 100 |

In the Alceste and IRaMuTeQ analyses, we utilized a lexicometric approach to extract insights from 275 articles published in various academic publications (Table 2). Reinert's conceptualization was the basis of our lexicometric research. The statistical treatments were carried out using Alceste software, with data derived from different databases to limit biases, including cross-sectional databases and functionally specialized databases (Chadegani et al., 2013).

Table 2. Sources of Database for Lexicometrics

| No. | Databases | N=135 | Disciplines |
|---|---|---|---|
| 1 | "IEEE express" | 15 | Technical, mechanics, engineering, telecommunication |
| 2 | "BSP premier" | 50 | Business science, economics, Law |
| 3 | "SCOPUS" | 50 | Business science, economics, Law |
| 4 | "AIS" | 20 | Management of information systems, operational research |

## 2. The Bibliometric Study

Bibliometric study aims to provide a comprehensive understanding of the development of blockchain management research and its current focus and direction. To achieve this goal, we conduct a bibliometric analysis that covers various aspects of the field. Firstly, we present a general overview of the research landscape, including the number of published papers, the countries of relevant journals, and the subject distribution. Then, we use co-citation and co-occurrence analyses to identify the research development path. Finally, we detect the burst of research hotspots and frontier citations. By combining and analysing existing literature, we hope to offer readers insights into the current state of blockchain management research and its future prospects.

### 2.1. Overview of Blockchain Management Research

Distribution of research disciplines: We conducted a subject category analysis to test the adequacy of multidisciplinary data using the 'Category' node in CiteSpace, as shown in Figure 2. This analysis examines the distribution of disciplines within our data set, which comprised 96 nodes and 374 lines, and yielded a Modularity Q value of 0.4925. Modularity Q rates the extent to which a network can be split into separate clusters. A network with low modularity cannot be subdivided into clusters with distinct borders, whereas a network with high modularity may be well-structured. Our analysis revealed that the network is significantly modularized, as the Modularity Q value exceeded 0.3 (Chen, 2010). Moreover, the Weighted Mean Silhouette (S) value, which measures the degree of uncertainty when evaluating the nature of a cluster, was 0.8474. This value indicates that the separation of clusters was efficient and convincing, as it exceeded the recommended threshold of 0.7 (Chen, 2006). Notably, the 'Computer Science' node was the most significant, with the largest tree ring history circle and a frequency of 330. We observed that computer science and engineering were the top nodes of the disciplines with higher frequencies, implying that research in technological infrastructure serves as the foundation for subsequent interdisciplinary applications. Meanwhile, the disciplines of business, economics, transportation, and automation systems had minor research scales. More than eight disciplines were developing their interactive innovation networks based on fundamental technical references, as indicated by the discipline distribution. Given the 2525 interdisciplinary data points, we proceeded with bibliometric research using co-citation analysis, timeline analysis, and noun-term

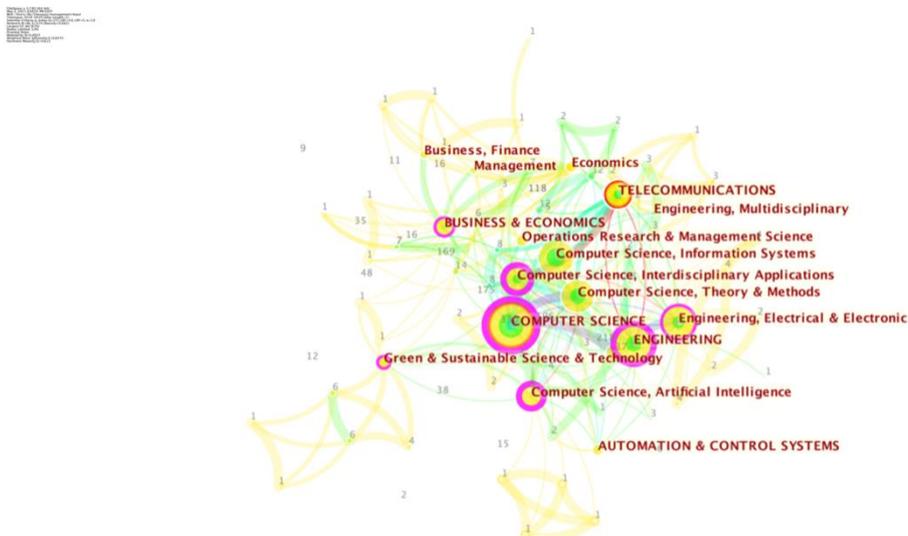

burst detection.

Figure 2. Discipline Distribution of Subject Category Analysis.

Countries of article origin: We analyse the countries that significantly contribute to the blockchain management database. The list considers not only citation frequency but also betweenness centrality. The betweenness centrality metric measures "the extent to which the node is in the middle of a path that connects other nodes in the network" (Freeman, 1977; Brandes, 2001). If a country's centrality values exceed 0.1, it will be considered to have revolutionary publications (Chen, 2005) and act as "gatekeepers" in social networks. In Table 3, we found that the USA, England, and China are the top three cited countries.

Although China has a quantitative advantage in citation frequency, its centrality is lower than the other two, indicating that blockchain management research in the USA is more centrally connected to the academic field. We were surprised to discover that France and Saudi Arabia exhibit high centrality despite having less than one hundred frequencies. Germany started researching blockchain management earlier than other countries, in 2014. While Canada and India also have high citation frequencies, their centrality metrics are below 0.1, meaning they cannot be identified as significantly influential.

Table 3. Citation Analysis of Countries

| No. | Freq | Centrality | Author | Publication Year |
|---|---|---|---|---|
| 1 | 426 | 0.2 | USA | 2015 |
| 2 | 187 | 0.18 | ENGLAND | 2016 |
| 3 | 716 | 0.17 | CHINA | 2015 |
| 4 | 66 | 0.15 | FRANCE | 2017 |
| 5 | 130 | 0.13 | AUSTRALIA | 2016 |
| 6 | 119 | 0.13 | GERMANY | 2014 |
| 7 | 123 | 0.11 | ITALY | 2016 |
| 8 | 80 | 0.1 | SAUDI ARABIA | 2018 |
| 9 | 113 | 0.09 | CANADA | 2016 |
| 10 | 221 | 0.06 | INDIA | 2017 |

## 2.2. Research Focus Analysis

Keyword co-occurrence analysis: According to the results of the co-occurrence analysis of the keywords shown in Table 4 below, the keywords 'blockchain' (1758), 'smart contract' (378), 'management' (339), and 'internet' (313) can be regarded as the major concepts in blockchain research. The trend in keyword usage suggests that 'blockchain' emerged as a technical term in 2014, while research interest in the management field began in 2016 and gradually increased due to the demand for supply chain solutions and the Ethereum boom in 2017. This indicates that blockchain research started from a technological infrastructure perspective and expanded to include practical applications and management, highlighting the maturity of the technology.

Table 4. List of the Keywords Frequency of Blockchain

| Publication Year | Count | Keywords |
|---|---|---|
| 2014 | 1758 | Blockchain |
| 2017 | 378 | Smart Contract |
| 2016 | 339 | Management |
| 2017 | 313 | Internet |
| 2017 | 278 | Security |
| 2017 | 197 | Internet of Things |

Co-citation analysis: We implemented co-citation analysis using CiteSpace. The software is equipped with seven algorithmic approaches: g-index, Top N, Top N%, Thresholds, Citations, Usage 180, and Usage 2013. We adopted the g-index to construct the co-citation network on 2525 articles (Zhu & Hua, 2017). Figure 3 displays the reference co-citation network from 2014 to 2021, where each color represents a year at the top of the figure. There are 685 nodes and 3337 lines. The tree ring history represents the highly cited nodes. We can see that citations show a burst in 2017 of the central light green. Christidis and Devetsikiotis (2016) is the most cited publication on blockchain, which proposed smart contracts in 2016. This contribution drove the evolution of blockchain from high volumes of Bitcoin research to IoT-related applications. Swan (2015) published a blueprint book to demonstrate the comprehensive development process of blockchain. Zheng et al. (2017) and Zyskind et al. (2015) from IEEE focus on the infrastructural architecture of consensus and decentralizing research. In recent years, we can see that the orange and red colors representing 2020 and 2021 gather on the upper right. Saberi et al. (2019) and Kshetri (2018) are studying the role of blockchain in supply chain management. Here we must clarify that the left purple node of Nakamoto, WOS database records Satoshi Nakamoto's work in 2019, which was published in 2008 (Nakamoto, 2008).

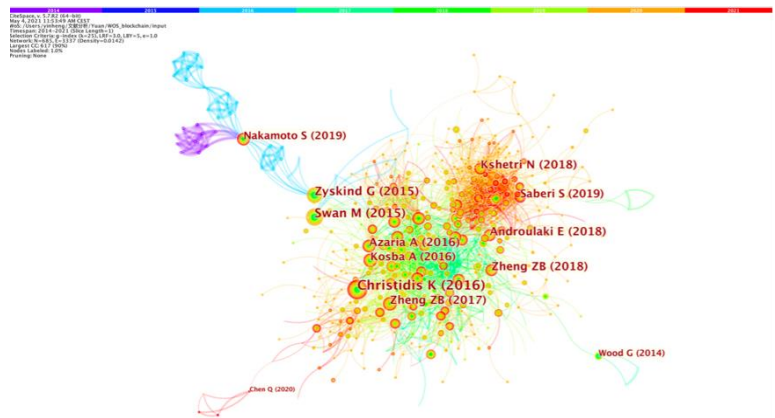

Figure 3. Map of the references of Blockchain

We also conducted a cluster analysis using Log-likelihood Ratio (LLR), which is shown in Figure 4. The analysis produced 12 clusters with a Modularity Q above 0.3 of 0.6115 and a Weighted Mean Silhouette above 0.7 of 0.8436. These values indicate that the analysis is a relatively significant and effective cluster network. In the early stages of 2014, the research topic focused on peer-to-peer infrastructure. From 2016 to 2019, the emphasis shifted towards practical applications of blockchain such as health records, mobile edge computing, and food supply chain. In the following two years, t there has been a greater focus on blockchain management issues, including capability-based access control and solutions. Therefore, our conclusion that blockchain is heading towards its management phase is consistent with these findings.

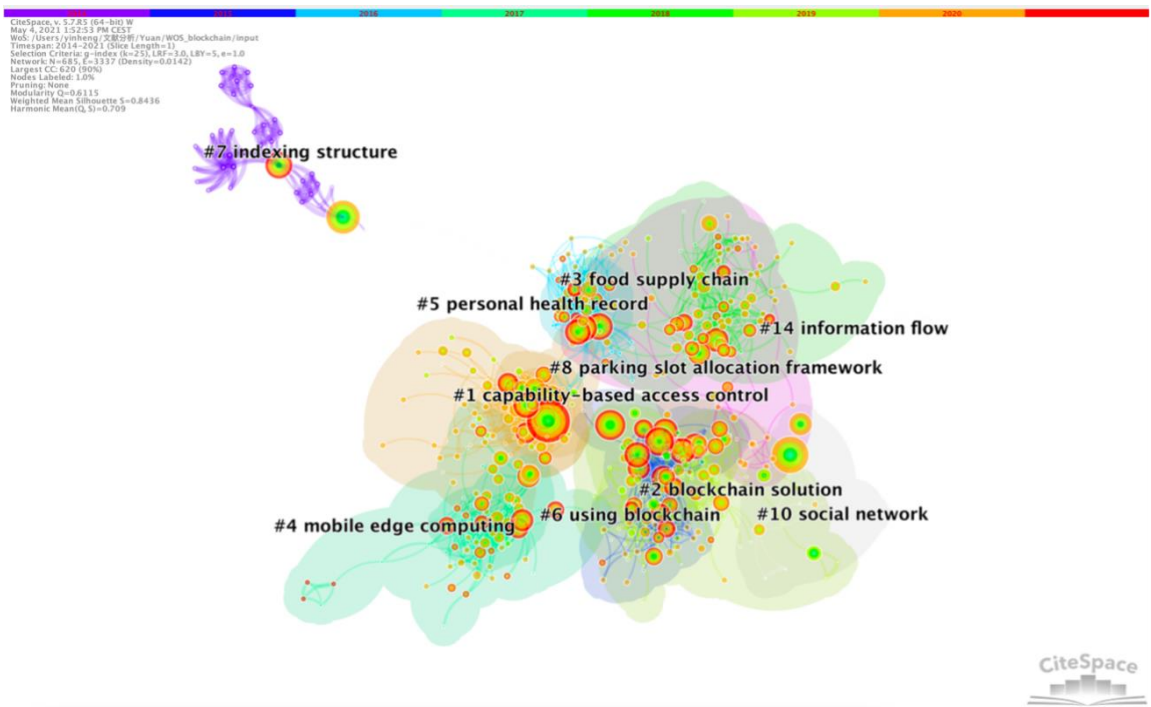

Figure 4. Map of the Clusters of Blockchain

Timeline visualizations: To investigate the developmental trajectory of blockchain, we used timeline visualizations to display the clusters and their links over time. The timeline analysis depicted in Figure 5 indicates that in the initial stages, blockchain research was primarily focused on system-building, which aligns with our previous visualization analysis. Blockchain then transitioned into an emergence phase in 2016 with the cloud computing federation and food supply chain integration. These developments broadened the research from technological engineering to interdisciplinary applications. However, it is noteworthy that cross-disciplinary expansion is still in a preliminary stage. In 2018, practical applications of mobile edge and supply chain management were successfully carried out. More nations and international corporations joined the research to develop and customize specific blockchain systems, including the national digital currency and Walmart food traceability system (Kamath, 2018; Bordo & Levin, 2017). The large scale of commercial use propels the blockchain research base forward. The similarity in structure and multiparty participation makes the supply chain a highly beneficial venture (Wang et al., 2019; Queiroz et al., 2020). Supply chains that synchronously update with blockchain applications have already made significant advancements in the management field (Beck et al., 2018). It is also a reminder that blockchain management is a multidisciplinary period.

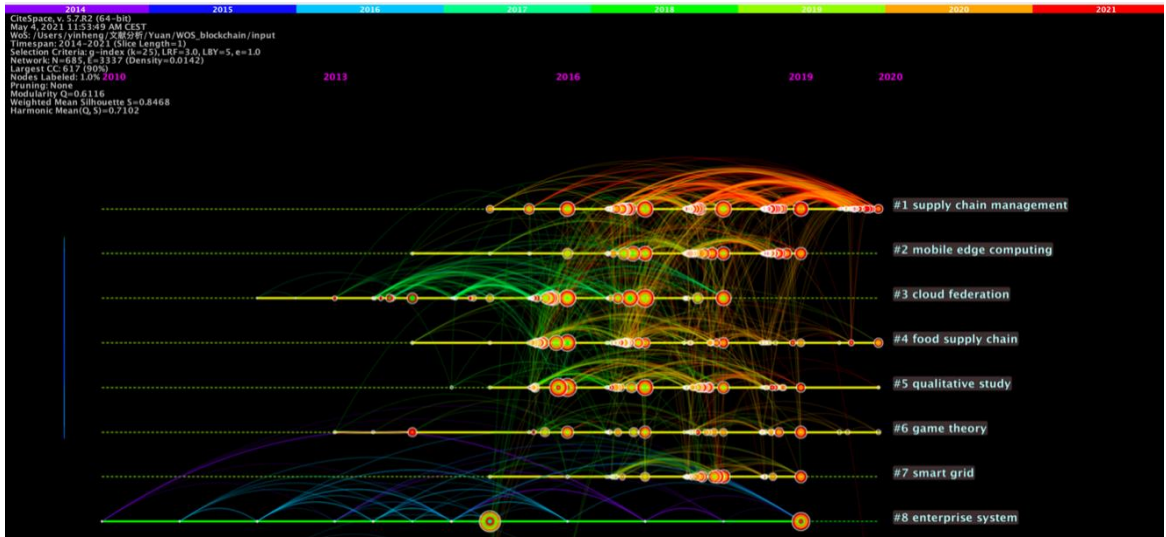

Figure 5. Map of the Timeline of Blockchain

## 2.3. Research Frontier Analysis of Blockchain Management

Analysis on Burst References: To gain further insight into the new research frontiers in the present co-cited references, we conducted an analysis of the cited bursts of the co-cited references. Burst detection is a method used to explore significant fluctuations during a short time interval. Figure 6 displays the top twenty-five references with the strongest citation bursts. These core articles are identified and listed in the same way as the keyword citation bursts. From the list, the earliest fundamental burst citation belongs to Nakamoto, which was published in 2008 but recorded late in 2019 in the WOS database. The strongest burst citation with a strength value of 20.22 belongs to Wood et al. (2014), who proposed the Ethereum Project in 2014. Additionally, Swan M and Zyskind G (2015), who were mentioned previously, generated new hotspots and attracted significant interest in the field of blockchain. These top publications play a crucial role in blockchain research because of their citation history, and they serve as a valuable review list with high-quality references for those interested in blockchain.

### Top 25 References with the Strongest Citation Bursts

| References | Year | Strength | Begin | End | 2014 - 2021 |
|---|---|---|---|---|---|
| Nakamoto S, 2019, BITCOIN PEER TO PEER, V0, P0 | 2019 | 7.27 | 2019 | 2018 | |
| Zyskind G, 2015, 2015 IEEE SECURITY AND PRIVACY WORKSHOPS (SPW), V0, P180, DOI | 2015 | 15.99 | 2016 | 2018 | |
| Swan M, 2015, BLOCKCHAIN BLUEPRINT, V0, P0 | 2015 | 13.26 | 2016 | 2018 | |
| Ali M, 2016, PROCEEDINGS OF USENIX ATC 16: 2016 USENIX ANNUAL TECHNICAL CONFERENCE, V0, P181 | 2016 | 6.3 | 2016 | 2018 | |
| Gubbi J, 2013, FUTURE GENER COMP SY, V29, P1645, DOI | 2013 | 4.75 | 2016 | 2018 | |
| Christidis K, 2016, IEEE ACCESS, V4, P2292, DOI | 2016 | 12.93 | 2017 | 2018 | |
| Wood G, 2014, ETHEREUM PROJECT YEL, V0, P0 | 2014 | 11.09 | 2017 | 2019 | |
| Kosba A, 2016, P IEEE S SECUR PRIV, V0, P839, DOI | 2016 | 9.65 | 2017 | 2018 | |
| Cachin C, 2016, WORKSH DISTR CRYPT C, V0, P0 | 2016 | 8.28 | 2017 | 2018 | |
| Narayanan A, 2016, BITCOIN CRYPTOCURREN, V0, P0 | 2016 | 7.17 | 2017 | 2018 | |
| Ben-Sasson E, 2014, P IEEE S SECUR PRIV, V0, P459, DOI | 2014 | 7.08 | 2017 | 2019 | |
| Bonneau J, 2015, P IEEE S SECUR PRIV, V0, P104, DOI | 2015 | 6.99 | 2017 | 2018 | |
| Garay J, 2015, PODC15: PROCEEDINGS OF THE 2015 ACM SYMPOSIUM ON PRINCIPLES OF DISTRIBUTED COMPUTING, V0, P0 | 2015 | 6.46 | 2017 | 2019 | |
| Buterin V, 2014, NEXT GENERATION SMAR, V0, P0 | 2014 | 6.15 | 2017 | 2019 | |
| Wright A, 2015, DECENTRALIZED BLOCKC, V0, P0 | 2015 | 5.97 | 2017 | 2018 | |
| Weber I, 2016, LECT NOTES COMPUT SC, V9850, P329, DOI | 2016 | 5.85 | 2017 | 2018 | |
| Antonopoulos AM, 2014, MASTERING BITCOIN UN, V0, P0 | 2014 | 5.54 | 2017 | 2019 | |
| Azaria A, 2016, PROCEEDINGS 2016 2ND INTERNATIONAL CONFERENCE ON OPEN AND BIG DATA - OBD 2016, V0, P25, DOI | 2016 | 4.67 | 2017 | 2018 | |
| Kraft D, 2016, PEER PEER NETW APPL, V9, P397, DOI | 2016 | 4.61 | 2017 | 2019 | |
| Wood G, 2014, ETHEREUM PROJECT YEL, V0, P1, DOI | 2014 | 20.22 | 2018 | 2019 | |
| Buterin V, 2014, ETHEREUM WHITE PAP, V3, P37 | 2014 | 8.45 | 2018 | 2019 | |
| Benet J, 2014, ARXIV PREPRINT ARXIV, V0, P0 | 2014 | 6.17 | 2018 | 2019 | |
| Croman K, 2016, LECT NOTES COMPUT SC, V9604, P106, DOI | 2016 | 5.45 | 2018 | 2019 | |
| Boudguiga A, 2017, 2017 2ND IEEE EUROPEAN SYMPOSIUM ON SECURITY AND PRIVACY WORKSHOPS (EUROS&PW), V0, P50, DOI | 2017 | 4.87 | 2018 | 2019 | |
| Ouaddah A, 2017, COMPUT NETW, V112, P237, DOI | 2017 | 4.87 | 2018 | 2019 | |

Figure 6. List of the Most Cited References of Blockchain

Trending topics: We conducted a keyword burst analysis using the same criteria settings as before. "Blockchain," "smart contract," and "management" remained the top three keywords according to frequency. However, there were differences between citation bursts and word frequency, as citation bursts indicate high interest in a topic during a specific period of time rather than total accumulated occurrences (Chen et al., 2010). Figure 7 shows the top thirteen keywords with the strongest citation bursts. The strength reflects the burst ratio to the time duration. The length of red lines represents the temporal mapping in a specific duration (Chen, 2006). Our analysis revealed that Bitcoin, which was collected from our research database in 2014, was the most relevant issue from 2015 to 2018. Ethereum and smart contracts, proposed in 2014, experienced bursts between 2017 and 2018, indicating the multi-field application and development of the technology. From 2018 onward, the research focus shifted to management schemes of distributed systems and applications. As theoretical designs have developed into business applications, social factors began to impact the relevance of higher-level research (Myers & Avison, 2002). Moreover, the practical application of technological studies has encroached on the socio-technical management domain, contributing to interdisciplinary research on the ex-post management of blockchain.

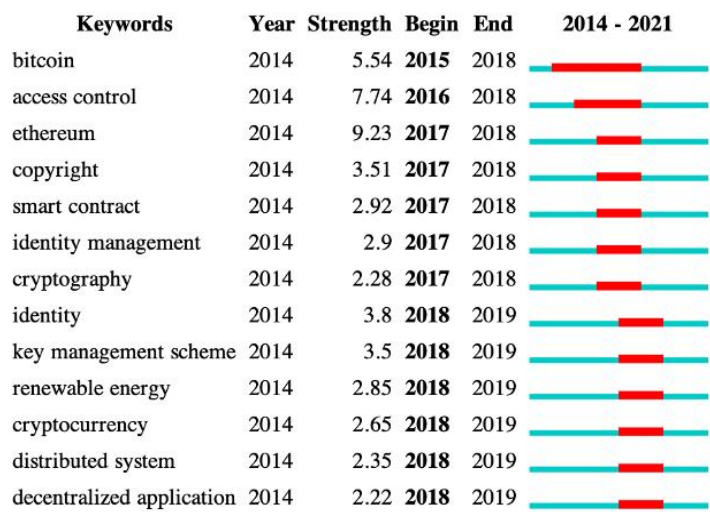

Figure 7 List of the Keywords Bursts of Blockchain

## 3. The Lexicometric Study

The bibliometric analysis reveals the extent of changes related to blockchain, but the lexicometric analysis helps to determine quantitatively whether those changes are disruptive. The advantage of analyzing textual data is the ability of deconstructing the corpus to simplify the material and access to the content of the corpus differently. The dataset detaches itself from the initial impressions of keyword appearances, bringing to light characteristics that might otherwise have gone unnoticed. This approach offers a more objective and rigorous examination of the content using statistical indicators, rather than relying on personal evaluations and first impressions. To further analyze textual data, the software initially lemmatizes the corpus by reducing verbs to their infinitive form, nouns to their singular form, and adjectives to their masculine singular form. Sections of the corpus are then analyzed using software such as Alceste and IRaMuTeQ to identify epistemic communities and extract meaning from the text.

The study identifies seven classes of words (as shown in the dendrogram in Figure 8 and Appendix), indicating differences in vocabulary used in work related to blockchains. The corpus consists of 572 text segments, containing 3511 different forms and 20549 occurrences. There are 2754 lemmas and 2392 active forms, with an additional 362 forms. The average number of forms per segment is 35.92. Out of 572 segments, 538 (94.06%) are classified. This dendrogram of the corpus allows us to identify lines of force (factors) between the lexical fields related to blockchain research. Each lexical world was entered using a list of forms ordered by decreasing chi2, which measures the significance of a form's presence in the class. Notably, the term "boundary" appeared in class 7, which relates to the epistemic community of production and logistics management.

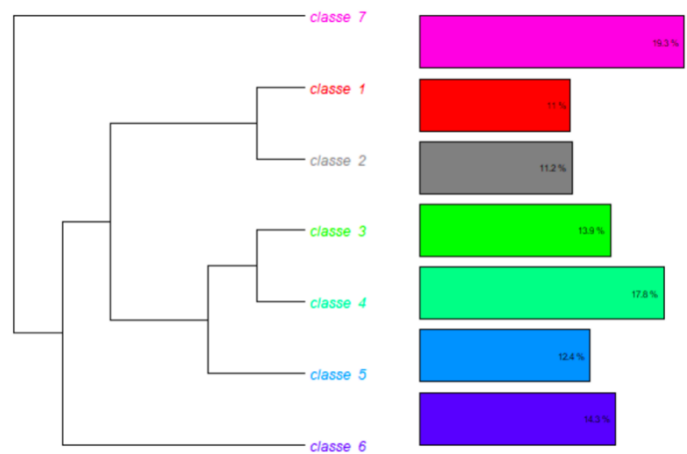

Figure 8. Factorial Analysis of the Correspondences

In the corpus, a prism of analysis class 3 refers to the link between IoT and blockchain. This class includes the words "community" and "architecture," which are associated with the epistemic community related to the internet of things. Unlike the internet revolution, which can be seen as an extension of a previous revolution, the blockchain revolution is expected to be cumulative and have corrective effects. As Zheng et al. (2018) suggest, blockchain has many advantages, including decentralization, persistence, anonymity, and auditability. Blockchain has a wide range of applications, including risk management of cryptocurrency financial services, internet of things, and public and social services (Bahga & Madisetti, 2016; Turk & Klinc, 2017; Banerjee et al., 2018; Ebrahimi et al., 2020). The incorporation of IoT has made objects intelligent and enabled them to communicate with each other, constantly capturing data from the physical world to analyse and perform intelligent actions. This has resulted in the accumulation of massive amounts of data, which is particularly relevant for logistics, transport, and industry, where the blockchain-IoT combination has the potential to significantly improve supply chain management and enhance efficiency.

## 4. Discussion

In order to assess whether blockchain represents a disruptive change, we will examine the implications of both bibliometric and lexicometric results. Specifically, we will analyse the stages of blockchain management based on bibliometric citations, as well as examine snapshots of the signifying structures extracted from quantified lexical texts.

### 4.1. Blockchain Management Is Becoming An Essential Issue

Our findings suggest that blockchain management and governance are critical issues for the ongoing development of the technology. The keyword bursts show that the hotspots trend from Bitcoin, Ethereum, smart contracts, decentralized applications, to management schemes. In addition, the concept of governance is matches problem-solving in blockchain's epistemic communities. This development process is similar to that of information systems and information system management, progressing from technology R&D, applications, and heading towards practical solutions and management. Furthermore, cryptographic assets are identified as intangible assets in GAAP and IFRS. Braune et al. (2020) stressed the importance of governance and financial structure in disclosing information about intangible capital. The governance of blockchain applications and their follow-up management is increasingly important.

### 4.2. The Shift from Technical Knowledge to Practical Knowledge

This study adds to the existing research on blockchains by demonstrating a transition from community of experts to communities of practice (CPs), which coincides with the maturation of the technology and the evolution from technical expertise to knowledge gained through practical applications. With the technology starting to have significant societal implications, the traditional centralized technology architecture is being replaced by multidisciplinary collaborations and interconnected systems. This shift is evident in the widespread adoption of the Internet of Things (IoT) in supply chain management to help with traceability, which reflects a shift from technical laboratory research to concrete, applicable tools.

### 4.3. Evolution Path of Blockchain Development

We have identified three distinct stages in the development of blockchain technology:

Emerging development phase (2008-2016): Blockchain was initially proposed in 2008 as an architectural design in computer science. During this period, blockchain was primarily focused on cryptocurrency as a peer-to-peer electronic cash system. In 2014, Ethereum introduced smart contracts, which made blockchain widely programmable and initiated application research in IoT, health records, finance, and supply chain from 2014-2016.

Technological upgrade phase (2016-2018): The introduction of smart contracts by Ethereum led to significant technological advancements and improved opportunities for interdisciplinary integration. As a result, blockchain solutions became foundational elements in back-end to develop decentralized applications. Each industry combined blockchain with its unique industrial characteristics, tackling diverse problems in different practical scenarios such as mobile edge computing, government services, supply chain, food safety, tax, and invoicing. This phase was characterized as a mutual upgrade process between underlying technology and its application.

Application-oriented phase (2018-Now): With the gradual maturation of blockchain technology, there has been a shift towards a more application-oriented phase. The concentrated focus on the managerial framework and scheme results in an efficient and user-friendly design. The integration of technology-specific fields leads to new research that builds upon practical concepts and considerations. The popularization of blockchain application promotes sustainable development and a dedicated management component.

### 4.4. Blockchain Is A Boundary Object From Epistemic Perspectives

Levina & Vaast (2005) and Ziolkowski et al. (2020) have studied the architecture of boundary objects in IT organizations, and identified the characteristics of blockchain using Cognitive and Hierarchical Models, as depicted in Table 5. From the perspective of knowledge and cognition, blockchain is currently perceived more as a boundary object than an element of disruptive change. This finding is also supported by the bibliometric study conducted by Tandon et al. (2021), which suggests that industry-based practitioners could benefit from delineating the research boundaries of individual areas of blockchain applications in various managerial domains, such as human resources, data management, and financial management. Their work implies that technological advancements require a broad understanding of blockchain's applicability across diverse sectors and managerial domains, concerning the management of business processes and operations (Tandon et al., 2021).

Modularity indicates a cognitive consideration of uncertainty (Azan et al., 2022). An object with multiple parts can be utilized in various situations, as highlighted by the lexicometric analysis that reveals the characteristic of multiple parts in a boundary object. However, blockchains cannot be considered a boundary object between CP, as they do not currently constitute either a boundary object in use or a boundary object in practice. Instead, it can be considered an epistemic boundary object, a technical object that mixes several types of knowledge and combines a few emerging practices without leading to knowledge capitalizations between practices. While blockchain technology is not only limited to CE, it also provides platforms for experimentation and simulation of future practices, as Saurel concluded in his oral intervention (2020). Some of these practices have already been mentioned in the literature (Chu, 2018; Cohney et al., 2019).

Table 5 Characteristics of Blockchain on Cognitive and Hierarchical Models

| Blockchain | Hierarchical model | Cognitive model |
|---|---|---|
| System | Representatives of the hierarchy (super-users, representative managers, etc.) | Cognitive representatives (up to the end users belonging to CPs) |
| Codification | Imposed by superiors and performed "in chambers." | Elaborated with the representatives of CPs |
| Boundary spanner | Nominated boundary spanners appointed by superiors | Boundary spanners in practice (legitimacy, trust, negotiation skills) |
| Boundary object | Designated boundary objects (designated by nominated boundary spanners by relying only on their own symbolic resources) | Boundary objects-in-use (local usefulness, common identity, versatility, abstraction, modularity, standardization) |
| Risks | Resistance to codification by CPs, non-use or diversion of tools, imposition of codification tools that destroy useful CPs for the company, etc. | Important investment of time and money |
| Advantages | Fast, low-cost | Reduces the risks of resistance or diversion |
| Knowledge transfer | Centralized, Top down, poor, efficient oriented, conflictual, creating resistant end-users | Emergent, practice-oriented, based on translation and on soft factors like uncertainty resistance, trust work motivation, trust. Enabled by participative design. |

## Conclusion

The research design provides an opportunity to measure the transformative power of blockchain on epistemic communities and academic knowledge. By utilizing quantitative visualization techniques, we can efficiently review blockchain studies and gain diverse perspectives on the blockchain knowledge domain (Yu & Sheng, 2020).

According to our analysis, blockchain research gained momentum in 2016, with management and governance research following in 2018. The USA, England, and China are the top three contributing nations with a higher relevance compared to others. We also identified the most influential researchers and the strongest citation bursts in academic research. The trending topics range from bitcoin, Ethereum, smart contracts to their management. The blockchain development path has been divided into three stages: emerging development, technological upgrade, and application-oriented use.

This paper contributes to blockchain research by demonstrating its boundary object characteristics due to integrated fusion of knowledge. The originality of this work lies in the combination of two kinds of algorithms to evidence the findings. However, there are some limitations to this study. For example, we cannot ascertain if the extraction from the literature will strictly correspond to the observed blockchain-based economic behaviour.

Our findings suggest that blockchain has no disruptive impact on scientific epistemic communities, although there is evidence of a very dynamic research trend. Blockchain research is not a simple technological development, but an interdisciplinary study that encompasses computer science, engineering, telecommunications, economics, and business management. Its theoretical framework comes from practical applications within communities. The results show that blockchain is currently still a boundary object, rather than a disruptive change from a knowledge perspective.

## Acknowledgment

This research was carried out in the frame of the cooperative doctoral programs (University of Lyon/CSC) supported by the China Scholarship Council (CSC). Yuan Li warmly acknowledges CSC for their support.

# Appendix

Table A. Classification of the corpus

| Class No. | Name of the class | Rate of significance |
|---|---|---|
| 1 | Monetary politics | 11% |
| 2 | Smart Contract | 11.2% |
| 3 | IoT & Communication | 13.9% |
| 4 | Distributed Ledger | 17.8% |
| 5 | Problem Solving | 12.4% |
| 6 | Governance | 14.3% |
| 7 | Production & SCM | 19.3% |